\documentclass[12pt]{article}
\usepackage{amsmath,amssymb,bm,graphicx}
\usepackage{color} 

\newcommand{\delslash}{\not \! \partial}

\begin{document}


\begin{center}
{\Large{\bf Seesaw mechanism and pseudo C-symmetry}}
\end{center}
\vskip .5 truecm
\begin{center}
{\bf { Kazuo Fujikawa$^{1}$ and Anca Tureanu$^2$}}
\end{center}

\begin{center}
\vspace*{0.4cm} 
{\it {$^1$Interdisciplinary Theoretical and Mathematical Sciences Program,\\
RIKEN, Wako 351-0198, Japan\\
\vspace*{0.2cm} 
$^2$Department of Physics, University of Helsinki, P.O.Box 64, 
\\FIN-00014 Helsinki,
Finland
}}
\end{center}
\vspace*{0.2cm} 
\begin{abstract}
It is shown that the specific "charge conjugation" transformation used to define the Majorana fermions in the conventional seesaw mechanism, namely $(\nu_{R})^{C}=C\overline{\nu_{R}}^{T}$ for a chiral fermion $\nu_{R}$ (and similarly for $\nu_{L}$), is a hidden symmetry associated with CP symmetry, and thus it formally holds independently of the P- and C-violating  terms in the CP invariant Lagrangian and it is in principle  applicable to charged leptons and quarks as well. This hidden symmetry, however, is not supported by a consistent unitary operator and thus it leads to mathematical (operatorial) ambiguities. When carefully examined, it also fails as a classical transformation law in  a Lorentz invariant field theory.
To distinguish it from the standard charge conjugation symmetry, we suggest for it the name of pseudo C-symmetry.  The pseudo C-symmetry is effective to identify Majorana neutrinos analogously to the classical Majorana condition. The analysis of CP breaking in weak interactions is performed using the conventional CP transformation, which is defined independently of the pseudo C-transformation, in the seesaw model after mass diagonalization.  
A way to ensure an operatorially consistent formulation of C-conjugation is to formulate the seesaw scheme by invoking a relativistic analogue of the Bogoliubov transformation.  

\end{abstract}
\section{Introduction}\label{intro}

Recent impressive developments in neutrino physics are well summarized in~\cite{particledata, xing,mohapatra2, fukugita, giunti, bilenky,valle}.
The main remaining issue is a better understanding of the extremely small neutrino masses, and the seesaw mechanism~\cite{minkowski,yanagida,mohapatra} provides a convenient framework to analyze this fundamental problem. 
The Lagrangian of the seesaw mechanism is left-right asymmetric, and thus the conventional parity is broken. If one assumes CP invariance, then the charge conjugation C is substantially broken. 
On the other hand, the exact solutions of the seesaw Lagrangian are the Majorana fermions that are the exact eigenstates of the charge conjugation by definition.
It is thus obvious that the charge conjugation to define the Majorana fermions in the seesaw mechanism cannot be identical to C, that defines CP and thus CPT of the starting seesaw Lagrangian.
  Moreover, the customary charge conjugation used to define the Majorana neutrino in the conventional seesaw scheme, when carefully examined,  leads to mathematical (operatorial) inconsistencies \cite{FT, FT2}. The purpose of the present paper is to clarify these puzzling aspects. 

In the following, we shall use the term {\it charge conjugation in seesaw} (and, later on, {\it pseudo C-transformation}) for the operation used in defining Majorana neutrinos in the seesaw scheme, and denote it by $\tilde C$. This operation is explained in more detail below. In contrast, we shall name {\it standard charge conjugation} and denote it by $C$, the usual operation of charge conjugation as is stated in standard textbooks on field theory~\cite{bjorken, weinberg1}.

The standard definitions of {\it classical} C, P and CP transformations for a Dirac field $\psi(x)$ are given by \cite{bjorken, weinberg1}
\begin{eqnarray}\label{C-P-CP1}
C&:& \psi(x) \rightarrow \psi^{C}(x)=C\overline{\psi}^{T}(x),\nonumber\\
P&:& \psi(x) \rightarrow i\gamma^{0}\psi(t,-\vec{x}),\nonumber\\
CP&:& \psi(x)\rightarrow i\gamma^{0}C\overline{\psi(t,-\vec{x})}^{T},
\end{eqnarray}
where we use the specific "$i\gamma^{0}$-parity" instead of the more common
$\gamma^{0}$-parity for the reasons stated in Section 2 of the paper. The charge conjugation matrix is $C=i\gamma^{2}\gamma^{0}$ in the convention of Ref. \cite{bjorken}.
The transformation laws for the chirally projected components of the Dirac field are defined 
by 
\begin{eqnarray}\label{C-P-CP2}
C&:& \psi_{L,R}(x) \rightarrow \psi^{C}_{L,R}(x)=C\overline{\psi_{R,L}}^{T}(x),\nonumber\\
P&:& \psi_{L,R}(x) \rightarrow i\gamma^{0}\psi_{R,L}(t,-\vec{x}),\ \ \psi^{C}_{L,R}(x) \rightarrow i\gamma^{0}\psi^{C}_{R,L}(t,-\vec{x}),\nonumber\\
CP&:& \psi_{L,R}(x)\rightarrow i\gamma^{0}C\overline{\psi_{L,R}(t,-\vec{x})}^{T}.
\end{eqnarray}
We recall that, if $\nu_{L}$ is a left-handed spinor, then $C\overline{\nu_{L}}^{T}$ is right-handed.  As C-conjugation as internal symmetry has to conserve chirality, a salient feature of these transformation laws is that we have the {\em doublet representations} for C and P, i.e. left- and right-handed fields are mixed. This is intuitively  easy to understand, because  charge conjugation is an internal transformation, taking particle to antiparticle, while chirality is a space-time property, identifying the $SU(2)$ subgroup of the Lorentz group $SO(1,3)=SU(2)_L\times SU(2)_R$ under which a certain two-spinor transforms nontrivially. On the other hand, we have a self-consistent transformation law for each chiral component in the case of CP 
symmetry. For these reasons, it is a well-known fact that, for an independent Weyl field, C and P transformations are undefined, while CP is well-defined in the same way as above.

Let us recall also that, in Lagrangian field theory,  we first define a {\it classical symmetry operation} and then look for  the {\it quantum operator} to realize it by Noether theorem in the case of continuous symmetries or other methods. In any quantum field theory, one should be able to define an operatorial realization of the charge conjugation transformation.
For a free Dirac quantum field $\psi(x)$, the unitary charge conjugation operation is defined as
\begin{eqnarray}\label{Dirac_cc_q}
\psi^{C}(x)\equiv{\cal C}\psi(x){\cal C}^{\dagger},
\end{eqnarray}
which acts on the creation and annihilation operators by changing the operators for particle into operators for antiparticle, without affecting their momentum and spin. The quantum operator ${\cal C}$ is realized in terms of creation and annihilation operator, according to a well-known prescription (see, e.g., \cite{bjorken}). Naturally, the classical and quantum charge conjugation operations have to coincide, i.e.
\begin{eqnarray}\label{Dirac_cc_c_q}
\psi^{C}(x)\equiv{\cal C}\psi(x){\cal C}^{\dagger}=C\overline{\psi}^{T}(x).
\end{eqnarray}
In conclusion, the charge conjugation transformation of a quantum Dirac field has both quantum and classical realizations and it mixes the left- and right-chirality components.

The conventional seesaw scheme~\cite{xing, fukugita, giunti, bilenky} constructs a Majorana fermion $\nu_{M}$ (which diagonalizes the mass term of the Lagrangian) from a chiral fermion $\nu_{R}$, for example, in the manner
\begin{eqnarray}\label{naive-Majorana}
\nu_{M}(x)=\nu_{R}(x)+\nu_{R}^{\tilde C}(x),
\end{eqnarray}
where
\begin{eqnarray}\label{naive-C}
\nu_{R}^{\tilde C}(x)\equiv C\overline{\nu_{R}}^{T}(x).
\end{eqnarray}
It is clear that the definition \eqref{naive-C} differs from what we would expect for the charge conjugation of a chiral component \eqref{C-P-CP2}. For this reason, we denote this operation by $\tilde C$ and we shall call it pseudo-C transformation (though in the literature it is denoted by $C$ and called C-conjugation proper).
It satisfies the relation
\begin{eqnarray}
\left(\nu_{R}^{\tilde C}\right)^{\tilde C}(x)=\nu_{R}(x)
\end{eqnarray}
and the Majorana-type condition
\begin{eqnarray}
\nu_{M}^{\tilde C}(x)=C\overline{\nu_{M}}^{T}(x)=\nu_{M}(x),
\end{eqnarray}
in other words, it satisfies at least two properties analogous to the standard classical charge conjugation. 

Our purpose is to analyze in depth this atypical charge conjugation concept and determine whether it is a sound notion in every respect. As we shall see below, the pseudo-C transformation does not respect: \\

i) the chirality conservation requirement;\\

ii) the operatorial realization requirement, in other words no quantum operator can be defined to implement the same transformation as \eqref{naive-C};\\

iii) internal consistency as a classical operation on spinors.\\

Let us analyze each point of the above checklist:\\

i)  It is well known that the pseudo-C conjugation, being defined for $\nu_{L}(x)$ and $\nu_{R}(x)$ {\em separately}, as in \eqref{naive-C}, changes the chirality of the field. The charge conjugation in seesaw is thus insensitive to the left-right mass asymmetry in the seesaw Lagrangian (see eq. \eqref{Lagrangian}).  This means that, were we able to find a quantum realization of it, that would change a particle of a given helicity to an antiparticle of the opposite helicity, therefore {\it pseudo-C could not be an internal transformation}.\\

ii) We {\it assume} the existence of a unitary operator ${\cal C}$ which satisfies
${\cal C}\nu_{R}(x){\cal C}^{\dagger}=\nu_{R}^{\tilde C}(x)$. Then,
\begin{eqnarray}\label{contradiction}
\nu_{R}^{\tilde C}(x)&=&{\cal C}\nu_{R}(x){\cal C}^{\dagger}\nonumber\\
&=&\frac{1+\gamma_{5}}{2}\,{\cal C}\nu_{R}(x){\cal C}^{\dagger}\nonumber\\
&=&\frac{1+\gamma_{5}}{2}\,C\overline{\nu_{R}(x)}^{T}=0
\end{eqnarray}
and similarly for $\nu_{L}(x)$. Here we used the fact that $\nu_{R}(x)=(\frac{1+\gamma_{5}}{2})\nu_{R}(x)$ and the left-handedness of $C\overline{\nu_{R}}^{T}(x)$ \footnote{Incidentally, for this definition of the seesaw charge conjugation operator, we formally have ${\cal C}\nu_{M}(x){\cal C}^{\dagger}={\cal C}\nu_{R}(x){\cal C}^{\dagger}+{\cal C}\left({\cal C}\nu_{R}(x){\cal C}^{\dagger}\right){\cal C}^{\dagger}
=C\overline{\nu_{R}(x)}^{T} + \nu_{R}(x) = \nu_{M}(x).$}. This sequence of equalities shows that there is a discrepancy between the classical definition of charge conjugation in seesaw and a possible quantum realization of it.\\
 
iii) One finds further puzzling aspects arising from the Ansatz \eqref{naive-C}. One can confirm that, using \eqref{naive-Majorana},
\begin{eqnarray}\label{free-Majorana}
S_{\rm Majorana}&=&\frac{1}{2}\int d^{4}x\,\overline{\nu_{M}}[i\delslash -m]\nu_{M}\nonumber\\
&=&\int d^{4}x \Big\{ \overline{\nu_{R}}i\delslash \nu_{R} -\frac{1}{2}\nu_{R}^{T}Cm\nu_{R} - \frac{1}{2}\overline{\nu_{R}}mC\overline{\nu_{R}}^{T} \Big\}\nonumber\\
&=&\int d^{4}x \Big\{\overline{\nu_{R}}i\delslash \frac{(1+\gamma_{5})}{2}\nu_{R}(x)-\frac{1}{2}\nu_{R}^{T}Cm\frac{(1+\gamma_{5})}{2}\nu_{R} + h.c. \Big\}.
\end{eqnarray}
If one assumes a  transformation rule of charge conjugation
in seesaw
\begin{equation}\label{naive_C_Classical}
\nu_{R}(x)\rightarrow \nu_{R}^{\tilde C}(x)= C\overline{\nu_{R}(x)}^{T},
\end{equation}
 as suggested by \eqref{naive-C}, it turns out that  the first and second expressions in \eqref{free-Majorana} are invariant under the transformation, while the last expression leads to a vanishing Lagrangian. We emphasize that the puzzling aspect in \eqref{free-Majorana} arises from the assumed classical transformation rule \eqref{naive_C_Classical}, irrespective of the existence or non-existence of the quantum  operator ${\cal C}$. Consequently, the example \eqref{free-Majorana} shows that even as a classical operation, the charge conjugation in seesaw \eqref{naive-C} is ambiguous. 

Remark that the three expressions in \eqref{free-Majorana} are identical as long as one assumes the relation \eqref{naive-Majorana}.  The reason for the vanishing of the last expression in
\eqref{free-Majorana} is that we use the  symmetry \eqref{naive-C} which is not consistently defined for {\em chiral fields}; the symmetry is not compatible with the explicit presence or absence of chiral projection operators $(1\pm\gamma_{5})/2$ in front of  chiral fields $\nu_{R,L}(x)$, namely, $[(1\pm\gamma_{5})/2]\nu_{R,L}(x)=\nu_{R,L}(x)$ in the Lagrangian.

Any sensible definition of parity reverses the chirality, and thus the CP transformation defined as  combination of the above pseudo-C conjugation \eqref{naive-C} and a suitable parity acts as $$\left((\nu_{L}(x))^{\tilde C}\right)^{P}\propto \gamma^{0}C\overline{\nu_{R}(t,-\vec{x})}^{T},$$ and thus cannot be a symmetry of weak interactions, for example. 

We thus see that the pseudo-C transformation \eqref{naive-C} fails on all three counts that we have listed as consistency checks for this C-conjugation notion. These are mathematical facts. On the other hand, when it is used in the conventional analysis of the seesaw Lagrangian, it appears to identify correctly the Majorana fermions. How comes then that an ill-defined concept formally leads to correct results? 
The basic technical reasons are that the operator condition \eqref{contradiction} has not been examined in practical applications and  the consistency of the last expression of the substitution rule \eqref{free-Majorana} has not been carefully checked in the past. 

The purpose of the present paper is to identify the theoretical origin of the pseudo-C conjugation \eqref{naive-C} which appears to work regardless of the formal violation of parity (left-right symmetry)  and thus the charge conjugation violation in the CP invariant seesaw model after the diagonalization  of the neutrino mass terms.  We shall show that the origin of the pseudo-C conjugation \eqref{naive-C} is in fact a hidden symmetry associated with  CP invariance, but the pseudo-C conjugation \eqref{naive-C} is derived from the CP symmetry  with an arbitrary truncation of the CP symmetry operation  in the seesaw model.  This explains the general validity of the pseudo-C conjugation \eqref{naive-C} in spite of the P and C violations in the CP invariant seesaw model and, at the same time, its technical difficulties in the operator condition \eqref{contradiction} and the last expression of the substitution rule \eqref{free-Majorana}. 
It will then be shown that a way to define the operatorially  consistent C-symmetry {and explain consistently how Majorana neutrinos are defined in the seesaw mechanism}  
is to use a relativistic analogue of the Bogoliubov transformation which has been formulated recently~\cite{FT,FT2, KF}. 

\section{Derivation of pseudo C-symmetry}\label{pseudo C}
\subsection{Seesaw Lagrangian}
We study a generic  Lagrangian for the three generations of neutrinos,
\begin{eqnarray}\label{Lagrangian}
{\cal L}&=&\overline{\nu}_{L}(x)i\gamma^{\mu}\partial_{\mu}\nu_{L}(x)+\overline{\nu}_{R}(x)i\gamma^{\mu}\partial_{\mu}\nu_{R}(x)\nonumber\\
&-&\overline{\nu}_{L}(x)m_{D} \nu_{R}(x)
-(1/2)\nu_{L}^{T}(x)C m_{L}\nu_{L}(x)\nonumber\\
&-&(1/2)\nu_{R}^{T}(x)C m_{R}\nu_{R}(x) + h.c.,
\end{eqnarray}
where $m_{D}$ is a complex $3\times 3$ Dirac mass matrix, and $m_{L}$ and $m_{R}$ are $3\times 3$ complex Majorana-type matrices. The anti-symmetry of the matrix $C$
 and Fermi statistics imply that $m_{L}$ and $m_{R}$ are symmetric. This is the Lagrangian of neutrinos with Dirac and Majorana mass terms. For $m_L=0$, it represents the classical seesaw Lagrangian of type I. In the following, we shall call the expression \eqref{Lagrangian} as the seesaw Lagrangian for the sake of generality.
 
We start with writing the mass term as 
\begin{eqnarray}
(-2){\cal L}_{mass}=
\left(\begin{array}{cc}
            \overline{\nu_{R}}&\overline{\nu_{R}^{C}}
            \end{array}\right)
\left(\begin{array}{cc}
            m_{R}& m_{D}\\
            m^{T}_{D}&m_{L}
            \end{array}\right)
            \left(\begin{array}{c}
            \nu_{L}^{C}\\
            \nu_{L}
            \end{array}\right) +h.c.,
\end{eqnarray}
where 
\begin{eqnarray}
\nu_{L}^{C}\equiv C\overline{\nu_{R}}^T, \ \ \ \nu_{R}^{C}\equiv C\overline{\nu_{L}}^T.  
\end{eqnarray}
Note that $\nu_{L}^{C}$ and $\nu_{R}^{C}$ are left-handed and right-handed, respectively. 
Since the mass matrix appearing is complex and symmetric, we can diagonalize it 
by a $6 \times 6$ unitary transformation $U$ (Autonne-Takagi factorization \cite{KF2}) as
\begin{eqnarray}\label{mass diagonalization}
            U^{T}
            \left(\begin{array}{cc}
            m_{R}& m_{D}\\
            m_{D}& m_{L}
            \end{array}\right)
            U
            =\left(\begin{array}{cc}
            M_{1}&0\\
            0&-M_{2}
            \end{array}\right)    ,        
\end{eqnarray}
where  $M_{1}$ and $M_{2}$ are $3\times 3$ real diagonal matrices (characteristic values). We choose one of the eigenvalues as $-M_{2}$ instead of $M_{2}$ since it is a natural choice in the case of a single generation model.

We thus have
\begin{eqnarray}\label{exact-mass}
(-2){\cal L}_{mass}
&=& \left(\begin{array}{cc}
            \overline{\tilde{\nu}_{R}}&\overline{\tilde{\nu}_{R}^{C}}
            \end{array}\right)
\left(\begin{array}{cc}
            M_{1}&0\\
            0&-M_{2} 
            \end{array}\right)            
            \left(\begin{array}{c}
            \tilde{\nu}_{L}^{C}\\
            \tilde{\nu}_{L}
            \end{array}\right) +h.c.,                       
\end{eqnarray}
with
\begin{eqnarray} \label{variable-change}          
            &&\left(\begin{array}{c}
            \nu_{L}^{C}\\
            \nu_{L}
            \end{array}\right)
            = U \left(\begin{array}{c}
            \tilde{\nu}_{L}^{C}\\
            \tilde{\nu}_{L}
            \end{array}\right)
           ,\ \ \ \ 
            \left(\begin{array}{c}
            \nu_{R}\\
            \nu_{R}^{C}
            \end{array}\right)
            = U^{\star} 
            \left(\begin{array}{c}
            \tilde{\nu}_{R}\\
            \tilde{\nu}_{R}^{C}
            \end{array}\right).          
\end{eqnarray}
Hence we can write 
\begin{eqnarray}\label{exact-solution}
{\cal L}
&=&(1/2)\{\overline{\tilde{\nu}_{L}}(x)i\delslash\tilde{\nu}_{L}(x)+ \overline{\tilde{\nu}_{L}^{C}}(x)i\delslash \tilde{\nu}_{L}^{C}(x)+\overline{\tilde{\nu}_{R}}(x)i\delslash \tilde{\nu}_{R}(x)\nonumber\\
&& \ \ \ \ \ + \overline{\tilde{\nu}_{R}^{C}}(x)i\delslash \tilde{\nu}_{R}^{C}(x)\}\nonumber\\
&-&(1/2)\left(\begin{array}{cc}
            \overline{\tilde{\nu}_{R}}&\overline{\tilde{\nu}_{R}^{C}}
            \end{array}\right)
\left(\begin{array}{cc}
            M_{1}&0\\
            0&-M_{2} 
            \end{array}\right)            
            \left(\begin{array}{c}
            \tilde{\nu}_{L}^{C}\\
            \tilde{\nu}_{L}
            \end{array}\right) +h.c.,
\end{eqnarray}
after transferring the possible CP violation contained in $U$ to the PMNS mixing matrix. 
In the present transformation \eqref{variable-change} in terms of a unitary matrix,  one can confirm that the conditions of canonical transformation
$ \tilde{\nu}_{L}^{C}=C\overline{\tilde{\nu}_{R}}^{T}$ and $
\tilde{\nu}_{R}^{C}=C\overline{\tilde{\nu}_{L}}^{T}$
hold after the transformation \cite{KF2}. 

 It is natural to define C, P and CP for the chiral variables in \eqref{exact-solution} following \eqref{C-P-CP2}. 
Among those symmetry transformations, the CP transformation 
\begin{eqnarray}\label{CP}
\tilde{\nu}_{L}(x)^{CP}=i\gamma^{0}C\overline{\tilde{\nu}_{L}(t,-\vec{x})}^{T}, \ \ \ \tilde{\nu}_{R}(x)^{CP}
=i\gamma^{0}C\overline{\tilde{\nu}_{R}(t,-\vec{x})}^{T}
\end{eqnarray}
is defined for theories only with $\tilde{\nu}_{L}$ or $\tilde{\nu}_{R}$.  In the present problem, we later confirm explicitly that the Lagrangian \eqref{exact-solution} is invariant under CP thus defined.

In the above definition of CP we adopted  the  transformation rule of ``$i\gamma^{0}$ parity'' which is defined, for a generic Dirac field, by
\begin{eqnarray}\label{conventional-P}
\psi^{P}(t,\vec{x})
=i\gamma^{0}\psi(t,-\vec{x}),
\end{eqnarray}
such that $\psi_{L,R}^{P}(t,\vec{x})
=i\gamma^{0}\psi_{R,L}(t,-\vec{x})$ which were used to infer the classical transformation laws of chiral fermions above.
The non-trivial phase freedom of the parity transformation in fermion number non-conserving theory has been analyzed by Weinberg \cite{weinberg1}. This definition of parity operation is the natural choice in a theory with Majorana fermions. The reason is that a  Majorana fermion $\psi_M(x)$, which is calssically defined by $\psi_{M}(x)(x)=C\overline{\psi_{M}}^{T}(x)$,  stays Majorana after parity transformation,  i.e. the parity transformation preserves the Majorana condition: $C\overline{i\gamma^{0}\psi_M(t,-\vec{x})} = i\gamma^{0}C\overline{\psi_M(t,-\vec{x})}=i\gamma^{0}\psi_M(t,-\vec{x})$ \cite{FT, FT2}.
The "$i\gamma^{0}$ parity" is crucial to assign a consistent intrinsic parity to an isolated Majorana fermion~\footnote{
In the full theory with charged leptons included, we assign the $i\gamma^{0}$-parity to charged leptons, for example, $e(x)\rightarrow i\gamma^{0}e(t,-\vec{x})$ for the sake of consistency, although the extra phase is cancelled in the lepton-number conserving terms.}.

The Lagrangian
\eqref{exact-solution} is then written in the form (by suppressing the tilde symbol for the chiral states $\tilde{\nu}_{R,L}$)
\begin{eqnarray}\label{exact-solution2}
{\cal L}
&=&(1/2)\{\overline{\psi_{+}}(x)i\delslash \psi_{+}(x)+ \overline{\psi_{-}}(x)i\delslash \psi_{-}(x)\}\nonumber\\
&-&(1/2)\{\overline{\psi_{+}}M_{1}\psi_{+}+ \overline{\psi_{-}}M_{2}\psi_{-}\}            
\end{eqnarray}
where 
\begin{eqnarray}\label{Majorana-variables3}
\psi_{+}(x)=\nu_{R}+ C\overline{\nu_{R}}^{T}, \ \ 
\psi_{-}(x)=\nu_{L}- C\overline{\nu_{L}}^{T}.
\end{eqnarray}
These variables satisfy the classical Majorana conditions
\begin{eqnarray} \label{identities}          
&&C\overline{\psi_{+}(x)}^{T}=C\overline{\nu_{R}}^{T}(x)+C\overline{C\overline{\nu_{R}}^{T}}^{T}(x)=C\overline{\nu_{R}}^{T}(x)+\nu_{R}(x)=\psi_{+}(x),\nonumber\\
&&C\overline{\psi_{-}(x)}^{T}=C\overline{\nu_{L}}^{T}(x)-C\overline{C\overline{\nu_{L}}^{T}}^{T}(x)=C\overline{\nu_{L}}^{T}(x)-\nu_{L}(x)=-\psi_{-}(x).
\end{eqnarray}
It is thus legitimate to look for some operator $\tilde{{\cal C}}$ which satisfies 
\begin{eqnarray} \label{pseudo C}          
&&\tilde{{\cal C}}\psi_{+}(x)\tilde{{\cal C}}^{\dagger}=\tilde{{\cal C}}\nu_{R}(x)\tilde{{\cal C}}^{\dagger}+\tilde{{\cal C}}C\overline{\nu_{R}}^{T}(x)\tilde{{\cal C}}^{\dagger}=C\overline{\nu_{R}}^{T}(x)+\nu_{R}(x)=\psi_{+}(x),\nonumber\\
&&\tilde{{\cal C}}\psi_{-}(x)\tilde{{\cal C}}^{\dagger}=\tilde{{\cal C}}\nu_{L}(x)\tilde{{\cal C}}^{\dagger}-\tilde{{\cal C}}C\overline{\nu_{L}}^{T}(x)\tilde{{\cal C}}^{\dagger}=C\overline{\nu_{L}}^{T}(x)-\nu_{L}(x)=-\psi_{-}(x).
\end{eqnarray}
From the comparison of \eqref{identities} and \eqref{pseudo C}, it may appear natural to guess that the operator $\tilde{\cal C}$ acts as follows:
\begin{eqnarray} \label{pseudo C operator}          
&&\tilde{{\cal C}}\nu_{R}(x)\tilde{{\cal C}}^{\dagger}=C\overline{\nu_{R}}^{T}(x),
\ \ \ \tilde{{\cal C}}C\overline{\nu_{R}}^{T}(x)\tilde{{\cal  C}}^{\dagger}=\nu_{R}(x),\nonumber\\
&&\tilde{{\cal C}}\nu_{L}(x)\tilde{{\cal C}}^{\dagger}=C\overline{\nu_{L}}^{T}(x), \ \ \ \tilde{{\cal C}}C\overline{\nu_{L}}^{T}(x)\tilde{{\cal C}}^{\dagger}=\nu_{L}(x).
\end{eqnarray}
This is precisely the pseudo C-symmetry transformation we discussed in \eqref{naive-C}. By the token of eq. \eqref{contradiction}, an operator $\tilde{\cal C}$ with the above properties cannot be defined.  

It is remarkable that the pseudo-C symmetry is {\it formally} an exact symmetry of the Lagrangian \eqref{exact-solution}, which is written as \eqref{exact-solution2}, but operatorially undefined.
We clarify the precise nature of the pseudo C-symmetry in the following.

We emphasize that the quadratic seesaw Lagrangian \eqref{Lagrangian} is brought by diagonalization to the form \eqref{exact-solution} or \eqref{exact-solution2} irrespective of the possible CP violation contained in the original parameters. By the PMNS matrix, after diagonalization, the CP violation is entirely transferred to the interaction terms. The pseudo C-transformation identifies the Majorana neutrinos in the {\it quadratic part of the Lagrangian}, therefore the discussion below is valid both for the CP-symmetric and CP-violating cases.

\subsection{Pseudo C-symmetry as hidden symmetry associated with CP invariance}
We shall examine now explicitly the CP-symmetry of the Lagrangian 
\eqref{exact-solution}, which is written by suppressing the tilde of the chiral variables, 
\begin{eqnarray}\label{exact-solution3}
{\cal L}
&=&\overline{\nu_{L}}(x)i\delslash\nu_{L}(x)+\overline{\nu_{R}}(x)i\delslash \nu_{R}(x)\nonumber\\
&-&(1/2)\{
   \nu_{R}^{T}CM_{1}\nu_{R} -\nu_{L}^{T}CM_{2}\nu_{L}
            +h.c.\}.
\end{eqnarray}
The analysis of CP invariance of the fermion number preserving terms in the Lagrangian is the usual one. We thus analyze the CP invariance of the fermion number violating terms:
\begin{eqnarray}
&&\int d^{4}x\, {\cal L}_{fermion-violating}^{CP}\nonumber\\
&=&\int d^{4}x[(1/2)\{i\gamma^{0}C\overline{\nu_{L}(t,-\vec{x})}^{T}\}^{T}M_{2}i\gamma^{0}C\overline{\nu_{L}(t,-\vec{x})}^{T}\nonumber\\
&-&(1/2)\{i\gamma^{0}C\overline{\nu_{R}(t,-\vec{x})}^{T}\}^{T}(x)C M_{1}i\gamma^{0}C\overline{\nu_{R}(t,-\vec{x})}^{T} + h.c.]\nonumber\\
&=&\int d^{4}x[(1/2)\{\overline{\nu_{L}(t,-\vec{x})}^{T}\}^{T}(x) M_{2}C\overline{\nu_{L}(t,-\vec{x})}^{T}\nonumber\\
&-&(1/2)\{\overline{\nu_{R}(t,-\vec{x})}^{T}\}^{T}(x) M_{1}C\overline{\nu_{R}(t,-\vec{x})}^{T} + h.c.]\nonumber\\
&=&\int d^{4}x[(1/2)\nu_{L}^{T}(x)C M_{2}\nu_{L}(x) - (1/2)\nu_{R}^{T}(x)CM_{1}\nu_{R}(x) + h.c.],
\end{eqnarray}
where we used $\{i\gamma^{0},C\}=0$ and $C^{T}C=1$ \cite{bjorken}. This shows the CP invariance, and we can promote the above CP transformation rule \eqref{CP} to a unitary operator in the context of the Lagrangian \eqref{exact-solution3}:
\begin{eqnarray}\label{operator-CP}
{\cal CP}\nu_{L}(x)({\cal CP})^{\dagger}
&=&i\gamma^{0}C\overline{\nu_{L}(t,-\vec{x})}^{T},\nonumber\\
{\cal CP}\nu_{R}(x)({\cal CP})^{\dagger}
&=&i\gamma^{0}C\overline{\nu_{R}(t,-\vec{x})}^{T}.
\end{eqnarray}

Now we come to the crucial observation of the present paper. We examine the CP transformation of the entire quadratic Lagrangian \eqref{exact-solution3}, but stop after cancelling the 
factor $i\gamma^{0}$ and changing the integration variables $-\vec{x}\rightarrow\vec{x}$. We then have 
\begin{eqnarray}
&&\int d^{4}x\,({\cal CP}){\cal L}(x)({\cal CP})^{\dagger}\nonumber\\
&=&\int d^{4}x[\overline{C\overline{\nu_{L}(x)}^{T}}i\gamma^{\mu}\partial_{\mu}C\overline{\nu_{L}(x)}^{T}+\overline{C\overline{\nu_{R}(x)}^{T}}i\gamma^{\mu}\partial_{\mu}C\overline{\nu_{R}(x)}^{T}\nonumber\\
&-&
(1/2)\{C\overline{\nu_{L}(x)}^{T}\}^{T}(x)C M_{2}C\overline{\nu_{L}(x)}^{T}\nonumber\\
&-&(1/2)\{C\overline{\nu_{R}(x)}^{T}\}^{T}(x)C M_{1}C\overline{\nu_{R}(x)}^{T} + h.c.]
\nonumber\\
&=&\int d^{4}x{\cal L}(x).
\end{eqnarray}
This relation shows a remarkable property, namely, the CP invariance of the quadratic seesaw Lagrangian implies that the action is formally invariant under the replacements
\begin{eqnarray}\label{pseudo-C2}
&&\nu_{L}(x)\rightarrow C\overline{\nu_{L}(x)}^{T},\nonumber\\
&&\nu_{R}(x)\rightarrow C\overline{\nu_{R}(x)}^{T},
\end{eqnarray}
independently of the values of mass parameters. Note that this symmetry is independent of space-time inversion, in spite of the fact that it changes the chirality of the field. This is precisely what we suggest to be called the
{\em pseudo C-symmetry}  \eqref{naive-C} of the seesaw Lagrangian. {\it A characteristic of the pseudo C-symmetry as a hidden symmetry associated with  CP invariance is that it is defined for any CP invariant theory even if  the separate well-defined P or C symmetries do not exist.}  Thus it is not influenced by the P and C violating left-right mass asymmetry of the seesaw Lagrangian. 

One can confirm that the relation \eqref{naive-Majorana} when $\nu_{M}(x)$ is treated as an independent field is
``covariant'' under CP symmetry up to the common factor $i\gamma^{0}$ on both sides together with spatial inversion $i\gamma^{0}\nu_{M}(t,-\vec{x})=i\gamma^{0}[\nu_{R}(t,-\vec{x})+\nu_{R}^{\tilde C}(t,-\vec{x})]$, while  the relation \eqref{naive-Majorana} is invariant under the pseudo C-symmetry without any spatial inversion.

The pseudo C-symmetry is very general but unfortunately {\em formal} as is seen in the identity, for example, 
\begin{eqnarray}\label{identity1}
&&\int d^{4}x \overline{\nu}_{L}(x)i\gamma^{\mu}\partial_{\mu}\nu_{L}(x)
=\int d^{4}x \overline{\nu}_{L}(x)i\gamma^{\mu}\partial_{\mu}\left(\frac{1-\gamma_{5}}{2}\right)\nu_{L}(x).
\end{eqnarray}
Both expressions in \eqref{identity1} are invariant under the CP symmetry \eqref{operator-CP}, while the first expression 
is invariant under the pseudo C-symmetry  \eqref{pseudo-C2} but the second expression vanishes under the same symmetry. It is important that the operatorial inconsistency  of the pseudo C-transformation in \eqref{contradiction} does not occur for the well-defined CP transformation in \eqref{operator-CP}. We thus understand the origin of the operatorial indefiniteness of the pseudo C-symmetry as arising from the arbitrary elimination of the factor $i\gamma^{0}$ of the CP transformation law \eqref{operator-CP} and thus resulting in the absence of a consistent unitary operator \eqref{contradiction} and the inconsistency in \eqref{free-Majorana}.

By generalizing the above argument, the pseudo C-transformation can be defined for any CP invariant quadratic Lagrangian and thus for the Standard Model, if one wishes. One can introduce the pseudo C-transformation for charged leptons and quarks also; for example, in the case of the electron it will read:  
\begin{eqnarray}\label{pseudo-C3}
&&e_{L}(x)\rightarrow C\overline{e_{L}(x)}^{T},\nonumber\\
&&e_{R}(x)\rightarrow C\overline{e_{R}(x)}^{T}.
\end{eqnarray}
In contrast, the standard C transformation is defined by $e_{L}(x)\rightarrow C\overline{e_{R}(x)}^{T}$ and $e_{R}(x)\rightarrow C\overline{e_{L}(x)}^{T}$.  
The CP invariant weak interaction Lagrangian (for a single generation model, for simplicity) is written as 
\begin{eqnarray}
 {\cal L}_{W}&=& (g/\sqrt{2})\overline{e_{L}}\gamma^{\mu}W^{-}_{\mu}\nu_{L} + h.c.\nonumber\\
 &=&(g/\sqrt{2})\overline{e_{L}}\left(\frac{1+\gamma_{5}}{2}\right)\gamma^{\mu}W^{-}_{\mu}\nu_{L} + h.c.
 \end{eqnarray}
One can confirm that the first expression in ${\cal L}_{W}$
is invariant under the pseudo C-symmetry \eqref{pseudo-C2} and \eqref{pseudo-C3} together with $W^{-}_{\mu}(x) \rightarrow W^{+}_{\mu}(x)$, while the second identical expression of ${\cal L}_{W}$ vanishes under the same pseudo C-symmetry~\cite{FT, FT2}. The pseudo C-symmetry is thus operatorially ill-defined.  This illustrates that an attempt to directly use the pseudo C-symmetry in neutrino phenomenology will be complicated.

The pseudo C-transformation changes the chirality, as compared to the ordinary C-conjugation and, to our knowledge,  a ``consistent CP symmetry'' defined as the combination of  the pseudo C-symmetry and a physically sensible parity operation, which may be used for weak interactions, has not been given.  In other words, the analysis of CP breaking in weak interactions has been performed using the CP transformation defined for chiral components in the seesaw model \eqref{exact-solution3} combined with the PMNS parameters (mixing angles and phase). The special property compared to the Standard Model without neutrino masses is that the neutrino number is not conserved as is indicated by the classical Majorana condition and also by the explicit form of the Lagrangian \eqref{exact-solution3}.

\section{Seesaw formulation with Bogoliubov transformation}\label{seesaw BT}

A way to avoid the use of the pseudo C-symmetry in the analysis of the seesaw model is to use the idea of a relativistic analogue of the Bogoliubov transformation \cite{FT, FT2}.
We illustrate the basic procedure by analyzing the single generation model for which we can work out everything explicitly by assuming CP invariant real mass parameters. 
We define a new Dirac-type variable
\begin{eqnarray}\label{4}
\nu(x)\equiv \nu_{L}(x) + \nu_{R}(x)
\end{eqnarray}
in terms of which the above Lagrangian is re-written as
\begin{eqnarray}\label{5}
{\cal L}&=&(1/2)\{\overline{\nu}(x)[i\delslash - m_D]\nu(x)+ \overline{\nu^{C}}(x)[i\delslash - m_D]\nu^{C}(x)\}\nonumber\\
&-&(\epsilon_{1}/4)[\overline{\nu^{C}}(x)\nu(x) +\overline{\nu}(x)\nu^{C}(x)]\nonumber\\
&-&(\epsilon_{5}/4)[\overline{\nu^{C}}(x)\gamma_{5}\nu(x) -\overline{\nu}(x)\gamma_{5}\nu^{C}(x)],
\end{eqnarray}
where $\epsilon_{1}=m_{R}+m_{L}$ and $\epsilon_{5}=m_{R}-m_{L}$.
The C and P transformation rules of $\nu(x)$ are defined by 
\begin{eqnarray}\label{7}
\nu^{C}(x)={\cal C}_{\nu}\nu(x){\cal C}_{\nu}^{\dagger}=C\bar{\nu}^{T}(x), \ \ \nu^{P}(x)={\cal P}\psi(x){\cal P}^{\dagger}=i\gamma^{0}\nu(t,-\vec{x}), 
\end{eqnarray}
and thus $\nu(x) \leftrightarrow \nu^{C}(x)$ under C and $\nu^{C}(x)\rightarrow i\gamma^{0}\nu^{C}(t,-\vec{x})$ under P; CP is given by 
$$\nu^{CP}(x)=i\gamma^{0}C\bar{\nu}^{T}(t,-\vec{x}).$$ 
The above  Lagrangian \eqref{5} is CP conserving, although C and P ($i\gamma^{0}$-parity) are separately broken by the last term for real $m_{D}$, $m_{L}$ and $m_{R}$. Note that here we are using the standard charge conjugation and parity transformation for Dirac fields. 
 
To solve \eqref{5}, we apply an analogue of Bogoliubov transformation between two sets of quantum fields, $(\nu, \nu^{C})\rightarrow (N, N^{C})$, defined as
\begin{eqnarray}\label{10}
\left(\begin{array}{c}
            N(x)\\
            N^{C}(x)
            \end{array}\right)
&=& \left(\begin{array}{c}
            \cos\theta\, \nu(x)-\gamma_{5}\sin\theta\, \nu^{C}(x)\\
            \cos\theta\, \nu^{C}(x)+\gamma_{5}\sin\theta\, \nu(x)
            \end{array}\right),
\end{eqnarray}
with
$\sin 2\theta =(\epsilon_{5}/2)/\sqrt{m_D^{2}+(\epsilon_{5}/2)^{2}}$.
We can then show that the anticommutators are preserved, i.e. $\{N(t,\vec{x}), N^{C}(t,\vec{y})\}=\{\nu(t,\vec{x}), \nu^{C}(t,\vec{y})\}$,
and thus it satisfies the canonicity condition of the Bogoliubov transformation. A transformation analogous to \eqref{10} is used in the analysis of neutron-antineutron oscillations \cite{FT}.

After the Bogoliubov transformation, which diagonalizes the Lagrangian with $\epsilon_{1}=0$,  ${\cal L}$ in \eqref{5} becomes
\begin{eqnarray}\label{13}
{\cal L}&=&\frac{1}{2}\left[\overline{N}(x)\left(i\delslash - M\right)
 N(x)+\overline{N^{C}}(x)\left(i\delslash - M\right)N^{C}(x)\right]\nonumber\\
&-&\frac{\epsilon_{1}}{4}[\overline{N^{C}}(x)N(x) + \overline{N}(x)N^{C}(x)],
\end{eqnarray}
with the mass parameter
\begin{eqnarray}\label{14}
M\equiv \sqrt{m_D^{2}+(\epsilon_{5}/2)^{2}}.
\end{eqnarray}
The Lagrangian \eqref{13} is invariant under the charge conjugation defined by $N^{C}(x)=C\overline{N(x)}^{T}$ and the $i\gamma^{0}$-parity defined by $N(x) \rightarrow i\gamma^{0}N(t,-\vec{x})$ and thus $N^{CP}(x) \rightarrow i\gamma^{0}N^{C}(t,-\vec{x})$.
The essence of the present Bogoliubov transformation is a CP-preserving canonical transformation which modifies the charge conjugation properties; for example, $\nu\leftrightarrow \nu^{C}$ does not lead to $N\leftrightarrow N^{C}$ in \eqref{10} in the operatorial sense, although the relation $N^{C}=C\overline{N}^{T}$ is maintained. It is crucial that C-noninvariant fermion number violating ``condensate'' with $\epsilon_{5}$ in \eqref{5} is converted to a C-invariant Dirac mass term of the Bogoliubov quasiparticle $N(x)$ in \eqref{13}.  A transformation to a theory of the Bogoliubov quasiparticle $N(x)$, which preserves both C and P, is a key to bypass the use of the pseudo C-symmetry. The parameter $\epsilon_{5}$ is an analogue of the energy gap in the BCS theory.

The Lagrangian \eqref{13} is exactly diagonalized by  
\begin{eqnarray}\psi_{+}(x)=\frac{1}{\sqrt{2}}(N(x)+N^{C}(x)),\cr
\psi_{-}(x)=\frac{1}{\sqrt{2}}(N(x)-N^{C}(x)),
\end{eqnarray}
in the form
\begin{eqnarray}\label{16}
{\cal L}&=&\frac{1}{2}\{\overline{\psi}_{+}[i\delslash-M_{+}]\psi_{+}
+\overline{\psi}_{-}[i\delslash-M_{-}]\psi_{-}\},
\end{eqnarray}
with masses $M_{\pm}=M \pm \epsilon_{1}/2$.
The charge conjugation and $i\gamma^{0}$-parity properties, which are induced by the transformation properties 
of $N(x)$, are  
\begin{eqnarray}\label{17}
&&\psi^{C}_{\pm}(x)=\pm \psi_{\pm}(x), \ \  \psi^{P}_{\pm}(x)= i\gamma^{0}\psi_{\pm}(t,-\vec{x}),
\end{eqnarray}
and thus define massive Majorana fermions.

It is straightforward to define the unitary charge conjugation operator  ${\cal C}_{M}$   for the free fermions $\psi_{\pm}(x)$, which satisfies
\begin{eqnarray}\label{Majorana cond seesaw}
{\cal C}_{M}\psi_{+}(x) {\cal C}^{\dagger}_{M}=C\overline{\psi_{+}(x)}^{T}=\psi_{+}(x),
\ \ \ \  {\cal C}_{M}\psi_{-}(x) {\cal C}^{\dagger}_{M}=C\overline{\psi_{-}(x)}^{T}=-\psi_{-}(x), 
\end{eqnarray}
with ${\cal C}_{M}|0\rangle_{M}=|0\rangle_{M}=|0\rangle_{N}$, following the procedure in the textbook \cite{bjorken}; in fact, the operator charge conjugation has the form ${\cal C}_{M}=\exp[i\pi \hat{n}_{\psi_{-}}]$, with the number operator $\hat{n}_{\psi_{-}}=\sum_{\vec{p},s}a^{\dagger}_{\psi_{-}}a_{\psi_{-}}$ of $\psi_{-}(x)$, and thus acts on $\psi_{+}(x)$ in a trivial manner.

The original neutrino is expressed in terms of the Majorana fermions $\psi_{\pm}$ if one uses \eqref{10} as 
\begin{eqnarray}\label{19}
\nu(x)&=&[(\cos\theta +\sin\theta \gamma_{5})/\sqrt{2}]\psi_{+}(x)+[(\cos\theta -\sin\theta \gamma_{5})/\sqrt{2}]\psi_{-}(x),
\end{eqnarray}
and  $\nu^{c}(x)=[(\cos\theta -\sin\theta \gamma_{5})/\sqrt{2}]\psi_{+}(x)-
[(\cos\theta +\sin\theta \gamma_{5})/\sqrt{2}]\psi_{-}(x)$, but 
the unitary C operations on $\psi_{\pm} \rightarrow \pm \psi_{\pm}$ in \eqref{19} do not reproduce $\nu^{c}(x)$, reflecting the C breaking in the original Lagrangian \eqref{5}.   

The Majorana fields $\psi_{\pm}(x)$ are the solutions of the exactly solvable Lagrangian \eqref{Lagrangian}. The vacuum defined by $\psi_{\pm}^{(+)}(x)|0\rangle_{M}=0$ is thus sufficient for all practical applications. But we encountered an analogue of Bogoliubov transformations, and thus it is interesting to examine the possible multiple-vacua structure. If one should define the vacuum by ${\cal C}_{\nu}(0)|0\rangle_\nu=|0\rangle_\nu$, then $|0\rangle_M \neq |0\rangle_\nu$, since one notes that ${\cal C}_{\nu}\neq {\cal C}_{M}$ if one defines ${\cal C}_{\nu}\nu(x){\cal C}^{\dagger}_{\nu}= \nu^{c}(x)$. In any case, ${\cal C}_{\nu}(t)$ is time-dependent since the C-transformation thus defined is not a symmetry of the Lagrangian \eqref{Lagrangian}. 
This implies that the vacuum of the Majorana fermions is different from the vacuum of the starting chiral fermions, when the latter are regarded as the chiral components of the Dirac neutrino field~\cite{FT,FT2, KF}~\footnote{In the  diagonalization of \eqref{Lagrangian}, one uses a unitary transformation $U$ of original chiral variables $\nu_{L,R}$ to mass eigenstates $\tilde{\nu}_{L,R}$ as in \eqref{variable-change}.
This transformation mixes the fermion and anti-fermion and in this sense changes the definition of the vacuum.}.

\section{Discussion and conclusion}

The conventional formulation of the seesaw mechanism~\cite{xing, fukugita, giunti, bilenky} customarily involves the use of a "pseudo C-symmetry" to define the Majorana fermions, to account for the phenomenological success of the seesaw mechanism. In this paper we have clarified the origin of this pseudo C-symmetry in the CP invariance of the quadratic seesaw Lagrangian. In principle, it is defined for any fermions such as charged leptons and quarks in the SM also.  The pseudo C-symmetry is thus very general, but it is operatorially undefined as we explicitly demonstrated, which is the mathematical fact.  
The CP symmetry breaking in weak interactions is analyzed using the CP symmetry of the chiral components appearing in the mass diagonalized seesaw Lagrangian \eqref{exact-solution3} combined with the PMNS mixing matrix.
The pseudo C-symmetry is effective to identify the Majorana neutrinos, although a similar identification is performed by the classical Majorana condition also.  It is hoped that the present analysis of the pseudo C-transformation will stimulate further analyses of this intriguing symmetry appearing in the mass generation of Majorana neutrinos.

The operatorial indefiniteness of the pseudo C-symmetry motivated us to reformulate the seesaw Lagrangian by rewriting it in a form analogous to the BCS theory~\cite{FT, FT2}. Then a relativistic analogue of Bogoliubov transformation leads to Majorana fermions in an
algebraically well-defined manner. The discrepancy between the C conjugation expected from the
original Lagrangian in the Dirac neutrino limit and the C conjugation in the picture of Majorana neutrinos is taken care of by an analogue of Bogoliubov transformation. The Bogoliubov transformation belongs to a class of canonical transformations which are more general than the familiar orthogonal or unitary transformations.
We emphasize that in this treatment, connecting $\nu_L$ and $\nu_R$ by charge conjugation and parity into a Dirac field as in \eqref{4} is a justified option and it is in the spirit of Bogoliubov's approach to the BCS theory \footnote{
Alternatively, we mention the use of a generalized Pauli--G\"ursey transformation. Pauli introduced the canonical transformation into neutrino physics.  The generalized Pauli--G\"ursey transformation based on group $U(6)$, which is a canonical transformation and mixes fermions and antifermions and thus induces multiple vacua with C defined on each vacuum, reproduces the result of our Bogoliubov transformation.  See \cite{KF2}.    }.
\\
\\
The present work was initiated when one of us (KF) was visiting Institute of High Energy Physics (IHEP), Chinese Academy of Sciences, Beijing. KF thanks Zhi-zhong Xing and Shun Zhou for  critical comments and hospitality at IHEP. We thank Masud Chaichian for helpful comments. KF is supported in part by JSPS KAKENHI (Grant No.18K03633).

\end{document}